\documentstyle[12pt]{article}

\baselineskip=\normalbaselineskip

\newcommand{\resection}[1]{\setcounter{equation}{0}\section{#1}}
\renewcommand{\theequation}{\thesection.\arabic{equation}}
\renewcommand{\thefootnote}{\fnsymbol{footnote}}
\newcommand{\bel}[1]{\begin{equation}\label{#1}}
\newcommand{\bal}[1]{\begin{eqnarray}\label{#1}}

\newcommand{\be}{\begin{equation}}
\newcommand{\ee}{\end{equation}}
\newcommand{\ba}{\begin{eqnarray}}
\newcommand{\ea}{\end{eqnarray}}
\newcommand{\nn}{\nonumber \\}
\newcommand{\qq}{\qquad}
\newcommand{\mat}[1]{\left(\matrix{#1}\right)}
\newcommand{\we}{\wedge}

\newcommand{\Lam}{\Lambda}

\newcommand{\n}{\nonumber}
\newcommand{\bra}[1]{\left\langle\,{#1}\,\right|}
\newcommand{\ket}[1]{\left|\,{#1}\,\right\rangle}
\newcommand{\bracket}[2]{
\left\langle\left.\,{#1}\,\right|\,{#2}\,\right\rangle}

\newcommand{\tr}{{\rm tr}}
\newcommand{\bR}{{\bf R}}
\newcommand{\bZ}{{\bf Z}}
\newcommand{\hg}{\widehat{g}}
\newcommand{\hphi}{\widehat{\phi}}
\newcommand{\hmu}{{\hat{\mu}}}
\newcommand{\hnu}{{\hat{\nu}}}
\newcommand{\hF}{{}F}
\newcommand{\hH}{\widehat{H}}
\newcommand{\hB}{\widehat{B}}
\newcommand{\Bn}[1]{B^{({#1})}}
\newcommand{\hBn}[1]{\widehat{B}^{({#1})}}

\newcommand{\An}[1]{A^{(1)\,{#1}}}
\newcommand{\bB}{{\bf B}}
\newcommand{\bpsi}{{\mbox{\boldmath{$\psi$}}}}
\newcommand{\sbpsi}{{\mbox{\boldmath{\scriptsize{$\psi$}}}}}
\newcommand{\bJ}{{\bf J}}
\newcommand{\bL}{{\bf \Lambda}}

\newcommand{\RR}{D}

\newcommand{\bRR}{{\bf \RR}}
\newcommand{\cE}{{\cal E}}
\newcommand{\bC}{{\bf C}}
\ifx\TwoupWrites\UnDeFiNeD\else\target{\magstepminus1}{11.3in}{8.27in}
\source{\magstep0}{7.5in}{11.69in}\fi
\newfont{\fourteencp}{cmcsc10 scaled\magstep2}
\newfont{\titlefont}{cmbx10 scaled\magstep3}
\newfont{\authorfont}{cmcsc10 scaled\magstep1}
\newfont{\fourteenmib}{cmmib10 scaled\magstep2}
\skewchar\fourteenmib='177
\newfont{\elevenmib}{cmmib10 scaled\magstephalf}
\skewchar\elevenmib='177
\makeatletter
\newcommand\nonsequentialeqnum{
\@addtoreset{equation}{section}
\def\theequation{\arabic{section}.\arabic{equation}}}
\newif\ifp@bblock \p@bblocktrue
\newcommand\nopubblock{\p@bblockfalse}
\newcommand\topspace{\hrule height 0pt depth 0pt \vskip}
\newcommand\p@bblock{\begingroup \tabskip=\hsize minus \hsize
\baselineskip=1.5\ht\strutbox \topspace-2\baselineskip
\halign to\hsize{\strut ##\hfil\tabskip=0pt\crcr
\the\Pubnum\crcr\the\date\crcr}\endgroup}
\renewcommand\titlepage{\ifx\TwoupWrites\UnDeFiNeD\null
\vspace{-1.7cm}\fi
\vskip0.6cm
\ifp@bblock\p@bblock \else\hrule height 0pt \relax \fi}
\makeatother
\newtoks\date
\newtoks\Pubnum
\newtoks\pubnum
\Pubnum={
~ \crcr
~ \crcr
YITP-99-43 \crcr
{\tt hep-th/9907132} \crcr
Revised Version
}
\newcommand{\frontpageskip}{\vspace{12pt plus .5fil minus 2pt}}
\renewcommand{\title}[1]{\frontpageskip
\begin{center}{\titlefont #1}\end{center}\par}
\renewcommand{\author}[1]{\frontpageskip\par\begin{center}
{\authorfont #1}\end{center}
\nobreak
}

\newcommand{\address}[1]{\par\begin{center}{\sl #1}\end{center}\par}

\renewcommand{\thanks}[1]{\footnote{#1}}
\renewcommand{\abstract}{\par\frontpageskip\centerline{
\fourteencp Abstract}
\vspace{8pt plus 3pt minus 3pt}}
\begin{document}

\titlepage

\renewcommand{\thefootnote}{\fnsymbol{footnote}}
\title{
\protect\Large{\protect\bf Comments on T-dualities \\
of Ramond-Ramond Potentials}
}

\author{
Masafumi Fukuma${}^{1\,}$\thanks{
e-mail address: {\tt fukuma@yukawa.kyoto-u.ac.jp}},
\,
Takeshi Oota${}^{2\,}$\thanks{
e-mail address: {\tt toota@tanashi.kek.jp}}\,
{{\rm and}}\,
Hirokazu Tanaka${}^{1\,}$\thanks{
e-mail address: {\tt hirokazu@yukawa.kyoto-u.ac.jp}}
}

\address{
${}^1$
Yukawa Institute for Theoretical Physics,\\
Kyoto University, Kyoto 606-8502, Japan \\
~\\
${}^2$
Institute of Particle and Nuclear Studies,\\
High Energy Accelerator Research Organization (KEK),\\
Tanashi, Tokyo 188-8501, Japan \\
}
\renewcommand{\thefootnote}{\arabic{footnote}}
\setcounter{footnote}{0}

%%%%%%%%%%%%%%%%%%%%%%%%%%%%%%%%%%%%%%%%%%%%%%%
% Abstract
%%%%%%%%%%%%%%%%%%%%%%%%%%%%%%%%%%%%%%%%%%%%%%
\begin{abstract}
The type IIA/IIB effective actions compactified on $T^d$ 
are known to be invariant under the T-duality group 
$SO(d, d; \bZ)$ 
although the invariance of the R-R sector is 
not so direct to see.
Inspired by a work of Brace, Morariu and Zumino,
we introduce new potentials which are mixture of
R-R potentials and the NS-NS 2-form in order to make 
the invariant structure of R-R sector more transparent.
We give a simple proof that if these new 
potentials transform as a Majorana-Weyl spinor
of $SO(d, d; \bZ)$, the effective actions are indeed 
invariant under the T-duality group.
The argument is made in such a way that 
it can apply to Kaluza-Klein forms of arbitrary degree. 
We also demonstrate that these new fields simplify 
all the expressions including the Chern-Simons term.

\vfil
\end{abstract}

%%%%%%%%%%%%%%%%%%%%%%%%%%%%%%%%%%%%%%%%%%%%%%%
% Main Part
%%%%%%%%%%%%%%%%%%%%%%%%%%%%%%%%%%%%%%%%%%%%%%%

%%%%%%%%%%%%%%%%%%%%%%%%%%%%%%%%%%%%%%%%%%%%%%%
%
\resection{Introduction}

Recent developments in string theory have been 
based on various kinds of duality symmetries. 
Among them, the T-duality was found first \cite{KY,SS}, 
which changes the size of the compactified space 
into its inverse in string unit. 
Although this symmetry was first recognized in the spectra 
of perturbative strings, it came to be believed  
that this should hold as an exact symmetry 
not simply as a perturbative one \cite{GPR}.
Later, at the level of low energy effective action,
the T-duality invariance of the type IIA/IIB theory was identified 
with a part of already known, much larger, 
and hidden symmetries of type II supergravities \cite{CSS}--\cite{CJ3}.
It was actually conjectured that the duality group of the full string 
theory can be extended to the U-duality group $E_{d+\!1 (d+\!1)}(\bZ)$ 
when compactified on a $d$-dimensional torus \cite{HT}.

Being a subgroup of the U-duality group, 
the T-duality group $SO(d,d;\bZ)$ has a special property: 
it is the maximum subgroup which consists of the elements 
that transform NS-NS and R-R fields into themselves, respectively. 
On the other hand, we sometimes encounter situations where NS-NS 
and R-R fields are better treated in a separate way. 
This is often the case when classical black-hole solutions 
of string theory are considered. 
Another example may be given by study of classical configurations 
based on the Born-Infeld action.
Thus, it should be useful if one can know in a simple manner 
how NS-NS and R-R fields transform under the T-duality group,
without resorting to embedding the whole structure 
once into the vast U-duality group.

The T-duality invariance can actually be seen very easily 
for the NS-NS sector of supergravity action \cite{MS}. 
There the kinetic term of the Kaluza-Klein (KK) scalars
$(G_{ij},B_{ij})$ $(i,j=1,...,d)$ can be written as\footnote{
This $B_{ij}$ will be denoted by $\Bn{0}_{ij}$ in the following sections 
to notify that this is a scalar for the noncompact 
$(10-d)$-dimensional space-time 
with coordinates $x^\mu~(\mu=0,1,...,9\!-\!d)$. 
We will also take the string unit $\alpha'=1$.} 
\ba
  {\cal L}_{\rm NS}=\frac{1}{8}\,
   e^{-2\phi}\,\tr\left(\partial_\mu M^{-1}\partial^\mu M\right)
\ea
with a $2d\times 2d$ matrix 
\ba
  M&=&\left( M_{rs} \right) \,=\,
   \mat{G^{-1} & -G^{-1}B \cr
     B G^{-1} & G-B G^{-1} B }
\ea
and the $(10\!-d)$-dimensional dilaton $\phi$. 
Thus the kinetic term is manifestly invariant under T-duality 
transformations $\Lambda\in O(d,d,\bZ)$ 
if the dilaton does not change 
and $M=(M_{rs})$ $(r,s=1,..,2d)$ transforms as 
\ba
    \overline{M}&=&\left(\Lambda^{-1}\right)^T\cdot M \cdot \Lambda^{-1}.
\ea
The KK 1-forms $(G_{\mu i},B_{\mu i})$ 
give a vector representation of $O(d,d;\bZ)$ 
and also have an invariant kinetic term \cite{MS}. 
These facts will be reviewed later in more detail.

On the other hand, the invariance of the sector 
including R-R potentials under the T-duality group 
$SO(d,d;\bZ)$\footnote{
Each of type IIA and type IIB is 
only invariant under the subgroup $SO(d, d; \bZ)$ of 
$O(d, d; \bZ)$, as we will see later.} 
is not so manifest as that for the NS-NS sector is.
There have actually been many works in which T-duality 
was studied as a subgroup of U-duality group
$E_{d+\!1 (d+\!1)}(\bZ)$ \cite{T}.
However, in order to write down the action in a manifestly 
U-invariant form, one needs to make a non-trivial mapping 
from the original fields to some other fields, 
which usually makes the T-duality symmetry 
for the original fields indirect. 
As for the works based on the T-duality itself, 
results have been obtained \cite{sugra} only for 
Nahm transformations which generate a subgroup of $O(d,d;\bZ)$.

By decomposing representations of 
$E_{d+\!1 (d+\!1)}(\bZ)$ with respect to $SO(d, d; \bZ)$,
it has been also known that Majorana-Weyl representations 
of $SO(d, d; \bZ)$ should appear in the R-R sector 
(see, for example, \cite{SV}).
However, as was discussed in detail for type IIA 
with $d=3$ in \cite{BMZ,BMZ2},
the R-R potentials themselves do not give Majorana-Weyl 
spinors directly. 
Instead, one needs to combine them 
with the NS-NS 2-form to get new fields that have 
such simple transformation properties
under $SO(d, d; \bZ)$.
Although prescription on how to arrange these fields
was known for each $d$ by starting from 11-dimensional supergravity 
\cite{SSz}, 
it is rather complicated due to the manner of field 
redefinitions which strongly depends on the dimensionality.
The main aim of this article is to present the prescription 
of constructing the new fields 
and to demonstrate the T-duality invariance of the R-R sector 
with the Chern-Simons term in a simple form.  
We give a discussion by investigating solely the structure of 
the effective action of type IIA/IIB strings 
with all fermionic fields set zero. 
Inclusion of fermions with analysis of supersymmetry 
will be discussed elsewhere. 
This work is inspired by analysis made 
by Brace, Morariu and Zumino \cite{BMZ,BMZ2}.

The main result can be summarized as follows. 
{}First, we introduce new potentials 
$\RR_{p+1}=(1/(p\!+\!1)!)\RR_{\hmu_1...\hmu_{p+1}} dx^{\hmu_1}
 \we \cdots \we dx^{\hmu_{p\!+\!1}}$ 
($\hmu_1,...,\hmu_{p+1}\!=\!0,1,...,9$)
which are mixtures of
R-R potentials and the NS-NS 2-form as
\ba
\begin{array}{lcl}
  \RR_0\,\equiv\,C_0, & \qq & \RR_1\,\equiv\,C_1, \\
  \RR_2\,\equiv\,C_2+\hB_2\wedge C_0, & &
                 \RR_3\,\equiv\,C_3+\hB_2\wedge C_1,\\
  \RR_4\,\equiv\,C_4+\frac{1}{2}\hB_2\wedge C_2
               +\frac{1}{2}\hB_2\wedge\hB_2\wedge C_0,
  & & ~
\end{array} 
\ea
where $C_{p+1}$ is the original $(p+1)$-form R-R potential 
and $\hB_2$ is the NS-NS 2-form in 10 dimensions.
We further introduce potentials of higher degree, 
$\RR_{p+1}$ $(p\!+\!1\!=\!5,...,8)$, as their electromagnetic duals. 
More precisely, we introduce the sum of field strengths 
\ba
{}  F\equiv e^{-\widehat{B}_2}\we\sum_{p+1=0}^8 d\RR_{p+1}
   = \sum_{p+2=1}^9\,F_{p+2},
\ea
and require the following relations in their equations of motion:
\ba
  \begin{array}{lll}
    \ast F_1=F_9, & \quad &\ast F_2=-F_8, \\
    \ast F_3=-F_7, & \quad & \ast F_4=F_6, \\
    \ast F_5=F_5, & \quad & \ast F_6=-F_4, \\
    \ast F_7=-F_3, & \quad & \ast F_8=F_2, \\
    \ast F_9=F_1. & \quad &
  \end{array}\label{const}
\ea
Note that $\ast^2\,F_n=(-1)^{n+1}F_n$ in 10-dimensional 
Minkowski space. 
The existence of these fields, $\RR_5,...,\RR_8$, 
is allowed by the equations of motion for 
$\RR_0,...,\RR_4$.

Our first claim is that, as far as the equations of motion 
are concerned, 
the R-R action with the Chern-Simons term 
can be rewritten into the following simple form: 
\ba
  S^{({\rm IIA})}_{\rm R+CS}&\equiv& \frac{1}{8\kappa_{10}^2}\,
   \int\,d^{10}x\,\sqrt{-\hg}\,
   \sum_{p+2=2,4,6,8}\,F_{p+2}\we \ast F_{p+2} \\
  S^{({\rm IIB})}_{\rm R+CS}&\equiv& \frac{1}{8\kappa_{10}^2}\,
   \int\,d^{10}x\,\sqrt{-\hg}\,
   \sum_{p+2=1,3,5,7,9}\,F_{p+2}\we \ast F_{p+2} \n
\ea
with all the $\RR_0,...,\RR_8$ being regarded as 
independent variables and with (\ref{const}) being 
the constraints to be imposed after the equations of motion 
are derived.

Second, for $d$-dimensional toroidal compactification, 
we assemble the set of KK scalars into the form 
$(\RR_\alpha)$ with $2^{d-1}$ entries:
\ba
\begin{array}{lcl}
 & d=1: & (\RR_\alpha)=(\RR_1) \\
 {\rm \underline{IIA:}}~~~~~~~~~~~ 
 & d=2: & (\RR_\alpha)=(\RR_1,\RR_2) \\
 & d=3: & (\RR_\alpha)=(\RR_1,\RR_2,\RR_3,\RR_{123}) \\
 & d=4: & (\RR_\alpha)=(\RR_1,\RR_2,\RR_3,\RR_4,
           \RR_{123},\RR_{124},\RR_{134},\RR_{234}) \\
 & \vdots & ~~~~~~~~\vdots 
\end{array}
\ea
\ba
\begin{array}{lcl}
 & d=1: & (\RR_\alpha)=(\RR) \\
 {\rm \underline{IIB:}}~~~~~~~~~~~ 
 & d=2: & (\RR_\alpha)=(\RR,\RR_{12}) \\
 & d=3: & (\RR_\alpha)=(\RR,\RR_{12},\RR_{13},\RR_{23}) \\
 & d=4: & (\RR_\alpha)=(\RR,\RR_{12},\RR_{13},\RR_{14},
            \RR_{23},\RR_{24},\RR_{34},\RR_{1234}) \\
 & \vdots & ~~~~~~~~\vdots
\end{array}
\ea
where $\RR_\alpha\!=\!\RR_{i_1...\,i_{p+1}}$ is the component of 
$\RR_{p+1}$ in the compact directions $y^{i_1},...,y^{i_{p+1}}$ 
$(1\!\leq\! i_1\! < \! \cdots \! < \! i_{p+1}\! \leq\! d)$.
Similarly, we also assemble the set of KK 1-forms 
$\RR_{\mu\,i_1...i_p}$ ($\mu=0,1,...,9\!-\!d$): 
\ba
\begin{array}{lcl}
 & d=1: & (\RR_{\mu\,\alpha})=(\RR_\mu) \\
 {\rm \underline{IIA:}}~~~~~~~~~~~ 
 & d=2: & (\RR_{\mu\,\alpha})=(\RR_{\mu},\RR_{\mu\,12}) \\
 & d=3: & (\RR_{\mu\,\alpha})=(\RR_{\mu},\RR_{\mu\,12},\RR_{\mu\,13},
      \RR_{\mu\,23}) \\
 & d=4: & (\RR_{\mu\,\alpha})=(\RR_\mu,\RR_{\mu\,12},\RR_{\mu\,13},
      \RR_{\mu\,14},\RR_{\mu\,23},\RR_{\mu\,24},\RR_{\mu\,34},
      \RR_{\mu\,1234}) \\
 & \vdots & ~~~~~~~~\vdots
\end{array}
\ea
\ba
\begin{array}{lcl}
 & d=1: & (\RR_{\mu\,\alpha})=(\RR_{\mu\,1}) \\
 {\rm \underline{IIB:}}~~~~~~~~~~~ 
 & d=2: & (\RR_{\mu\,\alpha})=(\RR_{\mu\,1},\RR_{\mu\,2}) \\
 & d=3: & (\RR_{\mu\,\alpha})=(\RR_{\mu\,1},\RR_{\mu\,2},
            \RR_{\mu\,3},\RR_{\mu\,123}) \\
 & d=4: & (\RR_{\mu\,\alpha})=(\RR_{\mu\,1},\RR_{\mu\,2},\RR_{\mu\,3},
            \RR_{\mu\,4},\RR_{\mu\,123},\RR_{\mu\,124},\RR_{\mu\,134},
            \RR_{\mu\,234}) \\
 & \vdots & ~~~~~~~~\vdots
\end{array}
\ea
This assembling may continue to KK forms of higher degree 
when $d$ is low enough.

Our second claim is that the dimensionally-reduced action 
of the R-R sector with the Chern-Simons term 
can be rewritten for type IIA and IIB, respectively, as\footnote{
Precise form is given by (\ref{main1})--(\ref{main3}).}
\ba
  {\cal L}_{\rm R+CS}\,=\,\frac{1}{4}\,\partial_\mu\!\RR_\alpha\, 
   S^\mp_{\alpha\beta}(M) \,\partial^\mu\!\RR_\beta,
  + \frac{1}{16}\,\partial_{[\mu}\RR_{\nu]\,\alpha}
   S^\pm_{\alpha\beta}(M) \,\partial^{[\mu}\!\RR^{\nu]}_{\,\,\beta} 
  + \cdots
  \label{main}
\ea
where $S^\pm_{\alpha\beta}(M)$ $(\alpha,\beta=1,...,2^{d-1})$ 
is a representation matrix of $M$ in the Majorana-Weyl
representation of $SO(d,d;\bR)$ with chirality $\pm$. 
The invariance of the action thus now becomes apparent 
by assuming that both of the $\RR_\alpha$ and $\RR_{\mu\,\alpha}$ 
transform as Majorana-Weyl spinors:
\ba
  \overline{\RR}_\alpha&=&S^\mp_{\alpha\beta}(\Lambda)\RR_\beta\\
  \overline{\RR}_{\mu\,\alpha}
   &=&S^\pm_{\alpha\beta}(\Lambda)\RR_{\mu\,\beta}.\n
\ea
We will prove the identity (\ref{main}) for arbitrary $d$ 
including KK forms of arbitrary degree. 
We simplify the argument with the use of the fermionic 
oscillator construction of Majorana spinor representation 
given in \cite{BMZ,BMZ2}.

The present paper is organized as follows. 
In section 2,  in order to fix our convention, 
we first give a brief review on the invariance 
of the NS-NS sector 
and then introduce new potentials $\RR_{p+1}$. 
In section 3, we explicitly construct the spinor representations 
of $O(d,d;\bZ)$ closely following
\cite{BMZ,BMZ2}, and then rewrite the R-R action plus 
the Chern-Simons term into a manifestly T-duality invariant form 
in section 4. 
Section 5 is devoted to discussions. 
The existence of the fields $\RR_5,...,\RR_8$ is proved in Appendix,  
with a demonstration that our new fields $\RR_{p+1}$ 
greatly simplify all the expressions including the Chern-Simons term.

%%%%%%%%%%%%%%%%%%%%%%%%%%%%%%%%%%%%%%%%%%%%%%%%%%%%%%%%%%%%%
%
\resection{Type IIA/IIB effective actions}

The action of ten-dimensional type IIA/IIB supergravity 
in the string metric can be split into three parts \cite{P}: 
\ba
  S=S_{\rm NS}+S_{\rm R}+S_{\rm CS}\label{actions}. 
\ea
The first term is the action for the NS-NS sector: 
\ba
  S_{\rm NS}=\frac{1}{2\kappa_{10}^2}\int d^{10}x\,\sqrt{-\widehat{g}}
    \,\,e^{-2\hphi}\left(\,\widehat{R}\,+\,4\,|d\hphi|^2_{\hg}
    \,-\,\frac{1}{2}\,|\widehat{H}_3|^2_{\hg}\,\right),\label{ns}
\ea
where $x^{\hmu}$ ($\hmu=0,1,...,9$) are 10-dimensional coordinates, 
and $\hg_{\hmu\hnu}$, $\hB_{\hmu\hnu}$ and $\hphi$ denote 
the 10-dimensional metric, NS-NS 2-form and dilaton, respectively. 
The NS-NS field strength is written as $\widehat{H}_3=d\hB_2$ 
with $\hB_2=(1/2)\hB_{\hmu\hnu}dx^{\hmu}\wedge dx^{\hnu}$. 
We adopt a rule that the subscript of a form stands for its degree 
when it has a definite meaning in 10 dimensions. 
We also often consider a sum of forms of various degree like 
$\Omega=\sum_K \Omega_K
=\sum_K(1/K!)\Omega_{\hmu_1...\hmu_K}dx^{\hmu_1}\wedge 
\cdots\wedge dx^{\hmu_K}$, 
and for this we introduce the invariant norm as 
\ba
  \left|\,\Omega\,\right|_{\widehat{g}}^2\,
    \equiv\,\sum_K\,\frac{1}{K!}\,\hg^{\hmu_1\hnu_1}
    \cdots\hg^{\hmu_K\hnu_K}\,
    \Omega_{\hmu_1...\hmu_K}\Omega_{\hnu_1...\hnu_K}.
\ea
The action for the R-R sector, $S_{\rm R}$, can be written 
for IIA and IIB, respectively, as 
\ba
  S^{({\rm IIA})}_{\rm R}
   &=&-\,\frac{1}{4\kappa_{10}^2}\int d^{10}x\sqrt{-\widehat{g}}
    \,\left(\,|\hF_{2}|^2_{\hg}\,+\,|\hF_{4}|^2_{\hg}\,\right)
    \label{ramond}\\
  S^{({\rm IIB})}_{\rm R}
   &=&-\,\frac{1}{4\kappa_{10}^2}\int d^{10}x\sqrt{-\widehat{g}}
    \,\left(\,|\hF_{1}|^2_{\hg}\,+\,|\hF_{3}|^2_{\hg}
      \,+\,\frac{1}{2}\,|\hF_{5}|^2_{\hg}\,\right),\n
\ea
where the R-R field strengths $\hF_{p+2}$ are defined from 
the $(p+1)$-form R-R potentials 
$C_{p+1}=(1/(p\!+\!1)!) C_{\hmu_1...\hmu_{p+1}}
dx^{\hmu_1}\!\we\!\cdots\!\we\! dx^{\hmu_{p+1}}$ 
as 
\ba
\begin{array}{lcl}
  \hF_1\,=\,dC_0, & \qq & \hF_2\,=\,dC_1, \\
  \hF_3\,=\,dC_2+\widehat{H}_3\wedge C_0, & &
                 \hF_4\,=\,dC_3+\widehat{H}_3\wedge C_1,\\
  \hF_5\,=\,dC_4+\frac{1}{2}\widehat{H}_3\wedge C_2
                  -\frac{1}{2}\,\hB_2\wedge dC_2. & & ~
\end{array} 
\ea
The Chern-Simons term $S_{{\rm CS}}$ is given by 
\ba
  S^{({\rm IIA})}_{\rm CS}&=&\frac{1}{4\kappa_{10}^2}\,\int\,
   \hB_2\we d C_3 \we d C_3 \label{cs}\\
  S^{({\rm IIB})}_{\rm CS}&=&\frac{1}{4\kappa_{10}^2}\,\int\,
   \hB_2\we d C_4 \we d C_2 .\n
\ea

We take a convention that NS-NS fields wear the hat 
($\widehat{~~}$) in 10 dimensions while R-R fields do not. 
This is because NS-NS fields generally need to be redefined 
after toroidal compactification in order to nicely behave
as fields living on the noncompact $(10-d)$-dimensional 
space-time (see, for example, (\ref{redef1}), (\ref{redef2}), 
(\ref{bhat}) and (\ref{redef3})).

After toroidal compactification on $T^d$, 
there will appear various KK forms both from 
the NS-NS and the R-R sectors.  
We first review the NS-NS case, closely following \cite{MS}.

\noindent\underline{\bf NS-NS sector:}

\noindent
We parametrize the 10-dimensional metric as
\ba
  d\widehat{s}^2&\equiv&\hg_{\hmu\hnu}dx^{\hmu} dx^{\hnu} \nn
  &=&g_{\mu\nu}dx^\mu dx^\nu 
    + G_{ij}(dy^i+A^i_\mu dx^\mu)(dy^j+A^j_\nu dx^\nu).\label{redef1}
\ea
Here the 10-dimensional coordinates are decomposed 
as $(x^{\hmu})=(x^\mu,y^i)$ $(\mu=0,1,...,9\!-\!d;~i=1,2,...,d)$, 
and we assume that all the fields depend only on 
the noncompact coordinates $x^\mu$.  
With this parametrization,  the kinetic term for potentials will 
take a complicated form since the KK 1-form
\ba
  \An{i}\equiv A^i_\mu dx^\mu 
\ea
will appear when contracting the indices in the compact directions. 
To simplify this, we follow the prescription of \cite{MS} 
which we found can be restated as follows. 
{}First, given a sum of forms 
$\Omega=\sum_K\Omega_K$, 
we decompose it as 
\ba
  \Omega = \sum_q\sum_n\frac{1}{n!}\,
    \Omega^{(q)}_{i_1...i_n}
    \,dy^{i_1}\wedge\cdots\wedge dy^{i_n},\label{organize}
\ea
where the superscript $(q)$ indicates that 
$\Omega^{(q)}_{i_1...i_n}$ is a $q$-form 
for noncompact indices: 
\ba
  \Omega^{(q)}_{i_1...i_n}=\frac{1}{q!}\,
     \Omega_{\mu_1...\mu_q\,i_1...i_n}
     dx^{\mu_1}\we\cdots\we dx^{\mu_q}. 
  \label{superscript}
\ea
Second, we introduce a new form $\Omega'$ by replacing 
$dy^i$ in $\Omega$ with $dy^i\!-\!\An{i}$, 
and reorganize it as in (\ref{organize}):
\ba
  \Omega'&\equiv& 
    \left.\Omega\,\right|_{dy^i\rightarrow\,dy^i\!-\!\An{i}}\nn
  &=&\sum_q\sum_n \frac{1}{n!}\,\Omega'^{\,(q)}_{i_1...i_n}\,
    dy^{i_1}\wedge\cdots\wedge dy^{i_n}. 
\ea 
Then the kinetic term can be expressed in such a way 
that all the indices are contracted only with 
$g^{\mu\nu}$ and $G^{ij}$:  
\ba
  \left|\,\Omega\,\right|_{\widehat{g}}^2
   =\left|\, \Omega'\,\right|^2_{g,G}
   \equiv\sum_q\sum_n \left|\,\Omega'^{\,(q)}_n \,\right|^2_{g,G}, 
\ea
where we have defined 
\ba
  \left|\, \Omega'^{\,(q)}_n \,\right|^2_{g,G}
   &\equiv&\frac{1}{n!}\,G^{i_1j_1}\cdots G^{i_nj_n}\,
    \frac{1}{q!}\,g^{\mu_1\nu_1}\cdots g^{\mu_q\nu_q}\,
    \Omega'_{\mu_1...\mu_q\,i_1...i_n}
    \Omega'_{\nu_1...\nu_q\,j_1...j_n}.
\ea
{}For example, the NS-NS field strength $\hH_3$ is rewritten as 
\ba
  \hH'^{\,(1)}_{ij}&=&d\Bn{0}_{ij}\nn
  \hH'^{\,(2)}_{i}&=&d\Bn{1}_{i}-\Bn{0}_{ij}d\An{j}\\
  \hH'^{\,(3)}    &=&d\Bn{2}-\frac{1}{2}\,\left(
     \Bn{1}_{i}d\An{i}+d\Bn{1}_i\An{i}\right), \n
\ea
where we have introduced 
\ba
  \Bn{0}_{ij}&\equiv&\hBn{0}_{ij} \nn
  \Bn{1}_{i}&\equiv&\hBn{1}_{i}+\hBn{0}_{ij}\An{j}\label{B2}\\ 
  \Bn{2}&\equiv&\hBn{2}-\frac{1}{2}\,\hBn{1}_{i}\An{i}.\n
\ea
Conversely, we have
\ba
  \hBn{0}_{ij}&\equiv&\Bn{0}_{ij} \nn
  \hBn{1}_{i}&\equiv&\Bn{1}_{i}-\Bn{0}_{ij}\An{j} \label{redef2}\\ 
  \hBn{2}&\equiv&\Bn{2}+\frac{1}{2}\,\Bn{1}_{i}\An{i}
        +\frac{1}{2}\,\Bn{0}_{ij}\An{i}\An{j},  \n
\ea
which give the original $\hB_2$ as 
\ba
  \hB_2&=&\frac{1}{2}\,\hBn{0}_{ij}dy^i\we dy^j
    +\hBn{1}_i dy^i + \hBn{2} \label{bhat}\\
  &=&\frac{1}{2}\,\Bn{0}_{ij}\left(dy^i+\An{i}\right)
      \left(dy^j+\An{j}\right)
    +\Bn{1}_i\left(dy^i+\An{i}\right)+\Bn{2}
    -\frac{1}{2}\,\Bn{1}_i\An{i}. \n 
\ea
Then the NS-NS part of the action will be rewritten \cite{MS} as
\ba
  S_{\rm NS}\,=\,\frac{1}{2\kappa_{10-d}^2}\int d^{10-d}x\sqrt{-g}\,
    {\cal L}_{\rm NS}, 
\ea
where
\be
\frac{1}{2\kappa_{10-d}^2} = \frac{1}{2\kappa_{10}^2}
\int d^d y.
\ee
By introducing the $(10-d)$-dimensional dilaton $\phi$ as 
\ba
  e^{-2\phi}\,\equiv\,e^{-2\hphi}\,\sqrt{G}, \label{redef3}
\ea
the factor ${\cal L}_{\rm NS}={\cal L}_1+{\cal L}_2+{\cal L}_3
+{\cal L}_4$ is given by 
\ba
  {\cal L}_1&=&e^{-2\phi}\left[
     R+4\,g^{\mu\nu}\partial_\mu\phi\,\partial_\nu\phi\right]\nn
  {\cal L}_2&=& \frac{1}{8}\,e^{-2\phi}\,g^{\mu\nu}\,\tr \left( 
    \partial_\mu M^{-1}\partial_\nu M\right)\nn
  {\cal L}_3&=&-\,\frac{1}{4}\, e^{-2\phi}\,
    g^{\mu_1\nu_1}g^{\mu_2\nu_2}
    {}F^{\,r}_{\mu_1\mu_2}\,M_{rs}\,F^{s}_{\nu_1\nu_2}\\
   {\cal L}_4&=& -\,\frac{1}{12}\,e^{-2\phi}\,
   g^{\mu_1\nu_1}g^{\mu_2\nu_2}g^{\mu_3\nu_3}
   \hH'_{\mu_1\mu_2\mu_3}\hH'_{\nu_1\nu_2\nu_3}\,,
  \n
\ea
where 
\ba
  &&\,M\,=\,\left( M_{rs} \right) \,\equiv\,
   \mat{G^{-1} & -G^{-1}\Bn{0} \cr
     \Bn{0} G^{-1} & G-\Bn{0} G^{-1} \Bn{0} },\qq
   \left( \Bn{0}\equiv \left(\Bn{0}_{ij}\right)\right), \nn
  &&\frac{1}{2}\,F^{\,r}_{\mu\nu}\,dx^\mu\wedge dx^\nu\,\equiv\,
    \left(\begin{array}{cc}
     d\Bn{1}_i \\ d\An{i} 
    \end{array}\right)\qq (r,s=1,...,2d;\,i,j=1,...,d).  
\ea
This form of action makes manifest its invariance under 
the T-duality group $O(d,d;\bZ)$ 
provided that the fields transform as 
\ba
  \overline{M}\,=\,\left(\Lambda^{-1}\right)^T
  \cdot M \cdot \Lambda^{-1}\,, \qq
  \left(\begin{array}{c}\overline{B}^{(1)}_i\\ \overline{A}^{(1)\,i}
    \end{array}\right)
    \,=\,\Lambda \left(\begin{array}{c}\Bn{1}_i\\
    \An{i}\end{array}\right)\,,\qq
  \overline{B}^{(2)}\,=\,\Bn{2}, 
\ea
for $\Lambda=\mat{a&b\cr c&d}\in O(d,d,\bZ)$ 
satisfying $\Lambda^T \, J \, \Lambda = J$ 
with $J=\mat{0 & 1_d \cr 1_d & 0}$.
The first transformation rule is equivalent to 
$\overline{E}\,=\,(aE+b)(cE+d)^{-1}$
for $E_{ij}=G_{ij}+\Bn{0}_{ij}$ \cite{GPR}.

\noindent\underline{\bf R-R sector with the Chern-Simons term:}

\noindent
The R-R potentials $C_{p+1}=(1/(p\!+\!1)!)\,
C_{\hmu_1...\hmu_{p+1}}\,dx^{\hmu_1}\wedge
\cdots\wedge dx^{\hmu_{p+1}}$ also produce  
KK forms of various degree after toroidal compactification.
To simplify all the expressions, 
we first combine the R-R potentials with the NS-NS 2-form 
in 10 dimensions as follows\footnote{
{}For the type IIA, the potentials $\RR_1$ and $\RR_3$ can be 
found in \cite{BMZ2}.}: 
\ba
\begin{array}{lcl}
  \RR_0\,\equiv\,C_0, & \qq & \RR_1\,\equiv\,C_1, \\
  \RR_2\,\equiv\,C_2+\hB_2\wedge C_0, & &
                 \RR_3\,\equiv\,C_3+\hB_2\wedge C_1,\\
  \RR_4\,\equiv\,C_4+\frac{1}{2}\hB_2\wedge C_2
             +\frac{1}{2}\hB_2\wedge\hB_2\wedge C_0.
  & & ~
\end{array} 
\label{newRR}
\ea
The R-R field strengths are then expressed with these $\RR_{p+1}$ as 
\ba
\begin{array}{lcl}
  \hF_1\,=\,d\RR_0, & \qq & \hF_2\,=\,d\RR_1, \\
  \hF_3\,=\,d\RR_2-\hB_2\wedge d\RR_0, & &
                 \hF_4\,=\,d\RR_3-\hB_2\wedge d\RR_1,\\
  \hF_5\,=\,d\RR_4-\hB_2\wedge d\RR_2
      +\frac{1}{2}\hB_2\wedge\hB_2\wedge d\RR_0. & & ~
\end{array} 
\ea 
These can be written in a simple form 
\ba
  \hF\,=\,e^{-\hB_2}\wedge d\RR \label{Fhat}
\ea
if we introduce 
\ba
  \RR\,\equiv\,\sum_{p+1=0}^4\RR_{p+1},\qq 
   \hF\,\equiv\,\sum_{p+2=1}^5\hF_{p+2}. 
\ea

The equations of motion for $\RR_0,...,\RR_4$ turn out 
to allow us to introduce extra R-R potentials of higher degree, 
$\RR_{p+1}$ $(p+1=5,...,8)$, that preserve the relation (\ref{Fhat}) 
with 
\ba
  \RR\,\equiv\,\sum_{p+1=0}^8\RR_{p+1},\qq 
   \hF\,\equiv\,\sum_{p+2=1}^9\hF_{p+2}, 
\ea
if we introduce the following identification for the field strengths 
of higher degree:
\ba
  \begin{array}{lll}
    \ast F_1=F_9, & \quad &\ast F_2=-F_8, \\
    \ast F_3=-F_7, & \quad & \ast F_4=F_6, \\
    \ast F_5=F_5, & \quad & \ast F_6=-F_4, \\
    \ast F_7=-F_3, & \quad & \ast F_8=F_2, \\
    \ast F_9=F_1 & \quad &
  \end{array}
  \label{elemag}
\ea
(see Appendix). 
Interestingly, as far as the equations of motion are concerned, 
we can in turn regard all the R-R potentials, $\RR_0,...,\RR_8$, 
as independent variables and choose 
\ba
  S^{({\rm IIA})}_{\rm R+CS}&\equiv&-\,\frac{1}{8\kappa_{10}^2}\,
   \int\,d^{10}x\,\sqrt{-\hg}\,
   \sum_{p+2=2,4,6,8}\,\left|F_{p+2}\right|^2_{\hg} 
   \label{newramond}\\
  S^{({\rm IIB})}_{\rm R+CS}&\equiv&-\,\frac{1}{8\kappa_{10}^2}\,
   \int\,d^{10}x\,\sqrt{-\hg}\,
   \sum_{p+2=1,3,5,7,9}\,\left|F_{p+2}\right|^2_{\hg}\n
\ea
as their action functional, 
with the understanding that the constraints (\ref{elemag}) are 
imposed after (and only after) the equations of motion are derived. 
In fact, one can prove that 
this system gives the same equations of motion 
with those from the sum of R-R and Chern-Simons terms 
$S_{\rm R}+S_{\rm CS}$, (\ref{ramond})--(\ref{cs}).  
We give a proof of this statement in Appendix.

{}For $d$-dimensional toroidal compactification, 
we introduce the primed field for $\hF$ as 
\ba
  {}F'&\equiv&\left.\hF\,\right|_{dy^i\rightarrow\,dy^i\!-\!\An{i}}.
  \label{Fprime}
\ea
Then the action for the R-R and Chern-Simons sector 
will be expressed as 
\ba
  S_{\rm R+CS}=\frac{1}{2\kappa_{10-d}^2}\,\int\,d^{10-d}x\,
   \sqrt{-g}\,{\cal L}_{\rm R+CS} \label{rcs1}
\ea
with 
\ba
  {\cal L}_{\rm R+CS}=-\frac{1}{4}\,
   \sqrt{G}\left|\,F'\,\right|^2_{g,G}. \label{rcs2}
\ea

To show that ${\cal L}_{\rm R+CS}$ is invariant under $O(d,d;\bZ)$ 
when the set of KK fields coming from $\RR$ transforms 
as a Majorana spinor of $O(d,d;\bZ)$,
in the next section we explicitly construct the spinor representation 
of $O(d,d;\bR)$ by using fermionic operators. 
We mostly follow the convention of \cite{BMZ,BMZ2}.

Before concluding this section, 
we would like to make a comment on the potentials $\RR_1$ and $\RR_3$ 
in the type IIA case.
It is well known that the type IIA
supergravity can be obtained from the 
$11$-dimensional supergravity \cite{CJ} by 
dimensional reduction. A coordinate transformation along
the $11$-th direction $x^{10} \rightarrow x^{10} + \xi$ 
becomes a $U(1)$ symmetry in $10$-dimensions:
\ba
\delta \hB_2 = 0, \qq
\delta C_1 = d\xi, \qq 
\delta C_3 = - \hB_2 \we d\xi.
\ea
Thus, these $\RR$ fields diagonalize 
the $U(1)$ symmetry :
$\RR_1 \rightarrow \RR_1+d\xi$, $\RR_3 \rightarrow \RR_3$.
These are $10$-dimensional analogues of
$A'$ fields of \cite{CJ2}.

%%%%%%%%%%%%%%%%%%%%%%%%%%%%%%%%%%%%%%%%%%%%%%%%%%%%%%%%%%%%%%
%
\resection{Spinor representation of $O(d, d;\bR)$}

We first recall that the group $O(d,d;\bR)$ 
consists of $2d \times 2d$ matrices $\Lambda$ satisfying
\be
  \Lambda^T \ J \ \Lambda= J, 
  \qq 
  J = 
  \mat{
    0 & 1_d \cr
   1_d & 0   }.
\ee
The group $O(d,d;\bZ)$ is defined as a subgroup 
that consists of matrices with integer-valued elements. 
It is known that both are generated by the following three types 
of matrices \cite{S}:
\ba
\Lambda_B &=& 
\mat{
1 & -B \cr
0 & 1  },
\qq
B^T = -B, \label{LamB} \\
\Lam_R &=& 
\mat{
R^{-1} & 0      \cr
0 & R^T },
\qq
R \in GL(d;\bR)~{\rm or}~GL(d;\bZ), \label{LamR} \\
\Lam_i &=&
-\mat{
1-e_i &  -e_i \cr
-e_i & 1-e_i },
\qq
(e_i)_{jk}=\delta_{ij} \delta_{ik},
\qq (i=1,...,d). \label{Lami}
\ea
Note that $\det \Lam_B=\det\Lam_R=+1$ and 
$\det\Lam_i=-1$. 
Thus one can construct a subgroup $SO(d,d;\bR)$ or 
$SO(d,d;\bZ)$ as such that are generated by 
$\Lam_B$, $\Lam_R$ and $\Lam_i\Lam_j$.

The Dirac matrices $\Gamma_r=(\Gamma_{r\,\alpha\beta})$ 
with $2^d\times 2^d$ components are introduced as 
\ba
  \left\{\Gamma_r,\,\Gamma_s\right\}=2J_{rs}\qq (r,s=1,...,2d), 
\ea
and the spinor representation 
$S(\Lam)=\left(S_{\alpha\beta}(\Lam)\right)$ 
is characterized by the property
\ba
  S(\Lam)\cdot\Gamma_s\cdot S(\Lam)^{-1}=\sum_r \Gamma_r 
   \,\Lam^{r}_{\,\,\,s}.\label{spinor}
\ea
To construct this representation, 
we introduce fermionic operators $\bpsi^{i\dag}$ and $\bpsi_i$ 
with the anti-commutation relations
\bel{metric}
\{ {\bpsi}_i, \bpsi^{j \dag} \} = {\delta_i}^j\,{\bf 1}, \qq
\{ \bpsi_i, \bpsi_j \} = 0 = \{ \bpsi^{i \dag}, \bpsi^{j \dag} \}
\qq
(i, j = 1,...,d). 
\ee
We define the hermitian conjugation as 
\be
(\bpsi_i)^{\dag} = \bpsi^{i\dag}, 
\ee
and introduce the vacuum $\ket{0}$ such that $\bpsi_i\ket{0}=0~
(i=1,...,d)$ and $\bracket{0}{0}=1$. 
Then the $2^d$-dimensional fermion Fock space 
is spanned by the vectors 
\ba
  \ket{\alpha}=
    \bpsi^{i_1\dag}\cdots\bpsi^{i_n\dag}\ket{0}\qq (n=0,...,d),
\ea
where $\alpha$ is a multi-index $\alpha=(i_1,...,i_n)~
(i_1\!\!<\!\!\cdots\!\!<\!\! i_n)$,  
and the Dirac matrices can be introduced with respect to this 
as follows: 
\ba
  \bpsi^{i\dag}\ket{\beta}&=&\sum_\alpha \ket{\alpha}\,
    \frac{1}{\sqrt{2}}\left(\Gamma_i\right)_{\alpha\beta}\nn
  \bpsi_i\ket{\beta}&=&\sum_\alpha \ket{\alpha}
    \frac{1}{\sqrt{2}}\left(\Gamma_{d+i}\right)_{\alpha\beta}. 
  \label{dirac}
\ea
Thus, if we can always introduce an operator $\bL$ to any element 
\ba
  \Lam = \mat{
    ({a_i}^j) & (b_{ij})  \cr
     (c^{ij}) & ({d^i}_j) } 
   \in O(d,d;\bR)
\ea
such that 
\ba
  \mat{
  \bL \bpsi^{j \dag} \bL^{-1},&
  \bL \bpsi_j \bL^{-1} } 
  &=& \mat{
  \bpsi^{i \dag} {a_i}^j + \bpsi_i c^{ij},&
  \bpsi^{i \dag} b_{ij} + \bpsi_i {d^i}_j } \nn
  &=& \mat{\bpsi^{i \dag},& \bpsi_i}
  \mat{
  a & b \cr
  c & d
  },\label{basic}
\ea
then, introducing the matrix $S_{\alpha\beta}(\Lam)$ by 
$\bL\ket{\beta}=\sum_\alpha\ket{\alpha} S_{\alpha\beta}(\Lam)$, 
we can establish the relation (\ref{spinor}). 
{}For this, it is enough to construct the operators that correspond to 
the elements given in (\ref{LamB})--(\ref{Lami}),
and it is easy to show that the followings are solutions 
\cite{BMZ,BMZ2}:
\ba
\bL_B 
   &=& e^{-\bB} \equiv \exp\left( -\frac{1}{2} B_{ij} \bpsi^{i\dag}
        \bpsi^{j\dag} \right), \nn
\bL_R &=& (\det{R})^{1/2} 
        \exp\left( -\bpsi^{i\dag} {A_i}^j \bpsi_j \right)
   \qq  \left(R=\left({R_i}^j\right) 
     = \exp\left({A_i}^j\right)\right), \nn
\bL_i &=& \bpsi_i + \bpsi^{i \dag} \qq (i=1,...,d). \label{hL}
\ea
Notice that all of these operators give real-valued matrix elements, 
so that the resulting representation is automatically Majorana. 
Note also that the $\bL_i$'s do not give a faithful representation 
so that there are always ambiguities in their orderings.

In order to construct Weyl representations, 
we define a matrix 
\ba
  \Gamma_{2d+1}\equiv\frac{1}{2^d}\,
    \prod_{i=1}^d\,(\Gamma_i+\Gamma_{d+i})(\Gamma_i-\Gamma_{d+i}) 
\ea
which satisfies $\left\{\Gamma_{2d+1},\Gamma_r\right\}=0~~
(r=1,...,2d)$. 
By looking at the correspondence (\ref{dirac}), 
one can easily see that $\Gamma_{2d+1}$ corresponds to 
$(-1)^{\bf N_F}$ with ${\bf N_F}=\sum_i\bpsi^{i\dag}\bpsi_i$. 
Thus, the projection to the subspace with $(-1)^{\bf N_F}=1$ 
leads to a Majorana-Weyl representation $\left(2^{d-1}\right)_s$ 
and the other one with $(-1)^{\bf N_F}=-1$ 
to $\left(2^{d-1}\right)_c$. 
Note that $\bL_i$ is a linear function of fermions 
and thus changes the chirality. 
Therefore, in order for an operator to preserve the chirality 
it must correspond to an element in $SO(d,d;\bR)$.

We further introduce an operator $\bJ$ that corresponds 
to $J=(J_{rs})$: 
\be
  \bJ = i^{d(d-1)/2} \ \bL_1 \cdots \bL_d,
\ee
where the phase factor is chosen such that $\bJ^2 = 1$.
One can actually prove that 
\ba 
  \bJ\ \bpsi^{i\dag}\ \bJ
   =\bpsi_i,\qq  \bJ\ \bpsi_i\ \bJ=\bpsi^{i\dag}.
\ea 
It is easy to check that for all these $\Lam$'s in (\ref{hL}) 
(and thus for all elements in $O(d,d;\bR)$), 
their transposes $\Lam^T=J\cdot\Lam^{-1}\cdot J$ are mapped to 
$\bL^{\dag}$:  
\be
  \bL^{\dag} = \bJ\bL^{-1}\bJ. \label{dagger}
\ee
In particular, we have
\be 
  \bL_B^{\dag} = e^{-\bB^{\dag}} = 
   \exp\left(\frac{1}{2} B_{ij}\, \bpsi_i \bpsi_j \right),\qq 
   \Lam_B^T = \mat{
     1 & 0 \cr
     B & 1 }.
\ee
Note also that the normalization of the operators (\ref{hL}) is 
correctly chosen such that they satisfy the condition (\ref{dagger}).

We finally make a comment that this operator $\bJ$ is 
essentially the charge conjugation operator. 
In fact, the operators defined by  
\ba
  \bC^{\pm} \equiv \bL^{\pm}_1 \cdots \bL^{\pm}_d 
\ea
with 
\ba
   \bL_i^{\pm}\equiv \bpsi^{i\dag}\pm \bpsi_i
\ea
can be easily seen to satisfy
\ba 
 \bC^\pm\,\left(\bC^\pm\right)^\dag={\bf 1}, &\qq&
    \left(\bC^\pm\right)^{\,2}=(-1)^{d(d\mp1)/2}\,{\bf 1},\\
 \bC^\pm\,\bpsi^{i\dag}\,\left(\bC^\pm\right)^{-1}
      =\mp\,(-1)^d\,\bpsi_i, &\qq&   
   \bC^\pm\,\bpsi_i\,\left({\bC^\pm}\right)^{-1}
      =\mp\,(-1)^d\,\bpsi^{i\dag}.   \n
\ea 
This implies that the matrices $C^\pm=(C^{\pm}_{\alpha\beta})$ 
defined by $\bC^\pm \ket{\beta}=\ket{\alpha}\,C^\pm_{\alpha\beta}$
satisfy the condition for the charge conjugation of $SO(d,d)$ 
\cite{KT}:
\ba
 && C^\pm\,\left(C^\pm\right)^\dag=1,\qq
  \left(C^\pm\right)^{\,T} = (-1)^{d(d\mp1)/2}\,C^\pm, \\
 && C^\pm\ \Gamma_r\ \left({C^\pm}\right)^{-1}
      =\mp\,(-1)^d\ \left(\Gamma_r\right)^T. \n
\ea

%%%%%%%%%%%%%%%%%%%%%%%%%%%%%%%%%%%%%%%%%%%%%%%%%%%%%%%%%%%%%%%%
%
\resection{R-R potentials and T-duality}

In this section, we show that the R-R action plus the Chern-Simons 
term after toroidal compactification on $T^d$, 
(\ref{rcs1})--(\ref{rcs2}), is actually invariant under $SO(d, d;\bZ)$ 
if a set of our R-R fields transform as a Majorana-Weyl spinor.

We first introduce a one-to-one correspondence between 
the set of forms and the space of creation operators 
by replacing the differential in the compact direction 
$dy^i$ with the fermion creation operator $\bpsi^{i\,\dag}$ 
as\footnote{
Recall that the superscript $(q)$ indicates that 
$\Omega^{(q)}_{i_1...i_n}$ is a $q$-form for noncompact indices 
(see (\ref{superscript})).}
\ba
  \Omega
   =\sum_n \frac{1}{n!}\,\,\Omega_{i_1...i_n}dy^{i_1}
      \we\cdots\we dy^{i_n}
   =\sum_q\sum_n \frac{1}{n!}\,\,
      \Omega^{(q)}_{i_1...i_n}dy^{i_1}\we\cdots\we dy^{i_n}
\ea
to
\ba
  {\bf \Omega}\equiv \sum_n \frac{1}{n!}\,\,\Omega_{i_1...i_n}
  \bpsi^{i_1\dag}\cdots \bpsi^{i_n\dag}
  =\sum_q \sum_n \frac{1}{n!}\,\,\Omega^{(q)}_{i_1...i_n}
  \bpsi^{i_1\dag}\cdots \bpsi^{i_n\dag}.
\ea
This actually gives an isomorphism as algebra. 
We also extend our rule such that $\Omega^{(q)}_{i_1...i_n}$ 
has ${\bf N_F}=q$, and thus it will anticommute with all the fermionic 
operators when $q$ is odd. 
We define a state corresponding to $\Omega$ as 
\ba
  \ket{\Omega}\equiv{\bf \Omega}\ket{0}.
\ea
Note that the following holds for any two forms $\Omega$ and $\Xi$: 
\ba
  {\bf \Omega}\ket{\Xi}=\ket{\Omega\we \Xi}.
\ea

Now that we have the above isomorphism, 
we can introduce the operator corresponding to $\hF$ in (\ref{Fhat}): 
\ba
  {\bf F}&=&e^{-\widehat{\bf B}_2}\,d\bRR .
\ea
Since $\hF_{p+2}$ are even (odd) forms for type IIA (IIB), 
we have  $(-1)^{{\bf N_F}}\ket{\hF}=+\ket{\hF}$ for type IIA 
and $=-\ket{\hF}$ for type IIB. 
This implies that each state has a definite chirality and 
thus forms a Majorana-Weyl representation of $SO(d,d;\bZ)$. 
Noticing that the replacement $dy^i\rightarrow dy^i\!-\!\An{i}$ 
as in (\ref{Fprime}) is equivalent 
to the operation 
\ba
  \bpsi^{i\dag}\rightarrow 
    e^{\sbpsi_i\An{i}}\bpsi^{i\dag}e^{-\sbpsi_i\An{i}}
    =\bpsi^{i\dag}\!-\!\An{i}, 
\ea
we can simply express the operator corresponding to $F'$ as 
\ba
  {\bf F'}=e^{\sbpsi_i\An{i}} {\bf F}e^{-\sbpsi_i\An{i}},
\ea
and thus the corresponding state can be written as 
\ba
  \ket{F'}&=&{\bf F'}\ket{0}=e^{\sbpsi_i\An{i}} {\bf F}\ket{0}\nn
    &=&e^{\sbpsi_i\An{i}}e^{-\widehat{\bf B}_2}\ket{d\RR}
    =e^{\sbpsi_i\An{i}}e^{-\widehat{\bf B}_2}e^{-\sbpsi_i\An{i}}
       \cdot e^{\sbpsi_i\An{i}}\ket{d\RR}.
\ea
Here one can use (\ref{bhat}) to show that 
\ba
  e^{\sbpsi_i\An{i}}\widehat{\bf B}_2 e^{-\sbpsi_i\An{i}}
    =\frac{1}{2}\,\Bn{0}_{ij}\bpsi^{i\dag}\bpsi^{j\dag}
    +\Bn{1}_i\bpsi^{i\dag}+\Bn{2}-\frac{1}{2}\,\Bn{1}_i\An{i}. 
\ea
Therefore we have 
\ba
  \ket{F'}&=&e^{-{\bf B}^{(0)}}e^{-\Bn{2}}e^{(1/2)\Bn{1}_i\An{i}}
    e^{\sbpsi^{i\dag}\Bn{1}_i}e^{\sbpsi_i\An{i}}\ket{d\RR}\nn
  &=&e^{-{\bf B}^{(0)}}e^{-\Bn{2}}e^{\bf V}\ket{d\RR},
  \label{Fprime2}
\ea
where 
\ba
  {\bf B}^{(0)}&\equiv&
      \frac{1}{2}\,B^{(0)}_{ij}\bpsi^{i\dag}\bpsi^{j\dag}\nn
  {\bf V}&\equiv& \bpsi^{i\dag}\Bn{1}_i+\bpsi_i\An{i}.
\ea
Since $\left(\Bn{1}_i,\,\An{i}\right)^T$ transforms as 
a vector of $O(d,d;\bZ)$, one can see that ${\bf V}$ transforms as 
\ba
  \overline{{\bf V}}=\bL \ {\bf V} \ \bL^{-1}
\ea
for $\Lam\in SO(d,d;\bZ)$. 
In fact, 
\ba
  \overline{{\bf V}}&=&(\bpsi^{\dag},\bpsi)
   \left( \begin{array}{c} 
         \overline{B}^{(1)}\\ \overline{A}^{(1)}
   \end{array}\right)
   =(\bpsi^{\dag},\bpsi)\ \Lam 
   \left(\begin{array}{c}
          B^{(1)}\\ A^{(1)}
   \end{array}\right)\nn
  &=&\bL\ (\bpsi^{\dag},\bpsi)\ \bL^{-1} 
   \left(\begin{array}{c}
          B^{(1)}\\ A^{(1)}
   \end{array}\right)
   =\bL\ {\bf V}\ \bL^{-1}. 
\ea

On the other hand, if we make a block-wise Gauss 
decomposition of $M$ as 
\ba
  M&=&\mat{1_d & 0 \cr B^{(0)} & 1_d}\cdot\mat{G^{-1}&0\cr0&G}\cdot
      \mat{1_d & -B^{(0)} \cr 0 & 1_d} \nn
  &=&\Lam_{B^{(0)}}^T \cdot \Lam_G\cdot  \Lam_{B^{(0)}}, 
\ea
then the corresponding operator ${\bf M}$ will be written as 
\ba
  {\bf M}=e^{-{\bf B}^{(0)}{}^{\dag}}\ \bL_G \ e^{-{\bf B}^{(0)}}
\ea
with 
\ba
  \bL_G\equiv \sqrt{G}\,e^{-\sbpsi^{i\dag}{h_i}^j\sbpsi_j} \qq
    \left(\left(G_{ij}\right)=e^{\left({h_i}^j\right)}\right).
\ea
This operator $\bL_G$ has a special property. 
In fact, suppose that for a given state  
\ba
  \ket{\Omega}=\sum_q\sum_n \frac{1}{n!}\,\Omega^{(q)}_{i_1...i_n}
    \,\bpsi^{i_1\dag}\cdots \bpsi^{i_n\dag}\ket{0},
\ea
we introduce its hermitian conjugate as 
\ba
  \bra{\Omega}=\sum_q\sum_n \frac{1}{n!}
    \,\bra{0}\bpsi_{i_n}\cdots \bpsi_{i_1}\,\ast_{10-d}
    \Omega^{(q)}_{i_1...i_n},
\ea
where $\ast_{10-d}$ is the Hodge-star in the noncompact 
$(10-d)$ dimensions. 
Then the following identity holds: 
\ba
  d^{10-d}x \sqrt{-g}\,\sqrt{G}\left|\,\Omega\,\right|^2_{g,G}
    =-\bra{\Omega} \bL_G \ket{\Omega}. \label{norm2}
\ea
In fact, using
\ba
  \bL_G \ket{\Omega}&=&\sum_q\sum_n\frac{\sqrt{G}}{n!}\,
   \left(e^{-h}\right)_{\!i_1}^{\,~j_1}\cdots
   \left(e^{-h}\right)_{\!i_n}^{\,~j_n}\Omega^{(q)}_{j_1...j_n}
   \bpsi^{i_1\dag}\cdots\bpsi^{i_n\dag}\ket{0}\nn
  &=&\sum_q\sum_n\frac{\sqrt{G}}{n!}\,
   G^{i_1j_1}\cdots G^{i_nj_n}\Omega^{(q)}_{j_1...j_n}
   \bpsi^{i_1\dag}\cdots\bpsi^{i_n\dag}\ket{0}, 
\ea
we can show 
\ba
  \bra{\Omega} \bL_G \ket{\Omega}&=&\sum_q \sum_n\,
   \frac{\sqrt{G}}{n!}\,
   \left(\ast_{10-d}\Omega^{(q)}_{i_1...i_n}
     \we\Omega^{(q)}_{j_1...j_n}\right)\,G^{i_1j_1}\cdots G^{i_nj_n}\\
  &=&-\,d^{10-d}x \sqrt{-g}\,\sqrt{G}\left|\,\Omega\,\right|^2_{g,G}.\n
\ea

Setting $\Omega=F'$ in (\ref{norm2}), we have 
\ba
    d^{10-d}x \sqrt{-g}\,\sqrt{G}\left|\,F'\,\right|^2_{g,G}
    =-\bra{F'} \bL_G \ket{F'}. 
\ea
Since this $F'$ is written as in (\ref{Fprime2}), 
the R-R action with the Chern-Simons term is expressed as 
\ba
  S_{\rm R+CS}&=&-\,\frac{1}{8\kappa_{10-d}^2}\,
   \int\,d^{10-d}x \sqrt{-g}\,\sqrt{G}\left|\,F'\,\right|^2_{g,G}\nn
  &=&\frac{1}{8\kappa_{10-d}^2}\,
   \int_{10-d}\,\bra{F'} \bL_G \ket{F'}\\
  &=&\frac{1}{8\kappa_{10-d}^2}\,
   \int_{10-d}\bra{K}{\bf M}\ket{K}\n
\ea
with
\ba
  \ket{K}=\exp\left(-B^{(2)}\right)\exp\left({\bf V}\right)
    \ket{d\RR}.
\ea
This can also be written as 
\ba
  S_{\rm R+CS}=\frac{1}{8\kappa_{10-d}^2}\,
   \int_{10-d} S_{\alpha\beta}(M)\,K_\alpha\we\ast_{10-d}K_\beta,
  \label{main1}
\ea
where $K_\alpha$ is a sum of forms in noncompact directions:
\ba
 K_\alpha=e^{-B^{(2)}}\we 
  \left(\displaystyle{e^{(1/\sqrt{2})\,\Gamma_r V^r}}\right)_{\alpha\beta}\we
  d\RR_\beta 
 \label{main2}
\ea
with
\ba
  B^{(2)}&=&\frac{1}{2}\,B_{\mu\nu}\,dx^\mu\we dx^\nu
    \quad ({\rm see~(\ref{B2})})\nn
  V^r&=&\left(
   \begin{array}{c}
    B_{\mu\,i}\,dx^\mu \\ A_{\mu}^i\,dx^\mu 
   \end{array}\right) \label{main3}\\
  \RR_\alpha&=&\sum_q\,\frac{1}{q!}\,dx^{\mu_1}\we\cdots\we 
   dx^{\mu_q}\,\RR_{\mu_1...\mu_q\,\alpha}\,.\n
\ea

Since ${\bf M}$, ${\bf V}$ and $B^{(2)}$ transform as 
\ba
  \overline{{\bf M}}
     =\left(\bL^{-1}\right)^{\dag} \ {\bf M} \ \bL^{-1},\qq
  \overline{{\bf V}}=\bL\ {\bf V}\ \bL^{-1},\qq
  \overline{B}^{(2)}=B^{(2)}, 
\ea
we see that 
the action is invariant under the whole T-duality group $SO(d,d;\bZ)$ 
if the $\RR=(\RR_\alpha)$ transforms as a Majorana-Weyl spinor:
\ba
  \ket{\overline{\RR}}=\bL \ket{\RR}. 
\ea
{}Furthermore, if we expand $\RR$ with respect to noncompact indices as 
\ba
  \RR=\sum_q\frac{1}{q!}\,dx^{\mu_1}\we\!\cdots\!\we dx^{\mu_q}\,
    \RR_{\mu_1...\mu_q}
\ea
with 
\ba
  \RR_{\mu_1...\mu_q}\equiv \sum_n\frac{1}{n!}\,
    \RR_{\mu_1...\mu_q\,i_1...i_n}\,dy^{i_1}\we\!\cdots\we dy^{i_n},
\ea
then each coefficient $\RR_{\mu_1...\mu_q}$ will also transform 
as a Majorana spinor:\footnote{
To be more precise, the following discussion holds 
only when $q+d\leq 10$.} 
\ba
  \ket{\overline{\RR}_{\mu_1...\mu_q}}=\bL \ket{\RR_{\mu_1...\mu_q}},
\ea
or equivalently,
\ba
  \overline{\RR}_{\mu_1...\mu_q\,\alpha}=\sum_\beta 
   S_{\alpha\beta}(\Lam)\,\RR_{\mu_1...\mu_q\,\beta}
\ea
with multi-indices $\alpha=(i_1,...,i_n)~~(i_1\!<\!\cdots\!<\! i_n;\,
n=0,...,d)$. 
Since $\RR_{\mu_1...\mu_q\,i_1...i_n}$ vanishes 
if $q+n={\rm even~(odd)}$ for type IIA (IIB), 
it has a definite chirality. 
This implies that 
$\RR_{\mu_1...\mu_q}=(\RR_{\mu_1...\mu_q\,\alpha})$ 
transforms as a Majorana-Weyl spinor 
for each set of noncompact indices $(\mu_1,...,\mu_q)$.

%%%%%%%%%%%%%%%%%%%%%%%%%%%%%%%%%%%%%%%%%%%%%%%%%%%%%%%%%%%%%%%
%
\resection{Discussion}

In the present article, we have given a simple proof 
that if the R-R potentials $C_{p+1}$ are combined with 
the NS-NS 2-form as in (\ref{newRR}), 
then their KK forms transform as Majorana-Weyl spinors 
under the T-duality group $SO(d,d;\bZ)$ 
in order to make the action invariant.

There should be various applications 
once transformation rules are obtained explicitly 
for the whole T-duality group. 
One will be to establish relations among various classical solutions 
of type IIA/IIB supergravities by using the full T-duality group 
together with the S-duality of type IIB. 
The work in this direction is in progress and 
will be reported elsewhere \cite{FOT}.

We finally make a comment on 
the dilaton dependence in R-R potentials, 
assuming the case $\hB_2=0$ in which 
there is no distinction between the original R-R potential 
$C_{p+1}$ and our potential $\RR_{p+1}$. 
Usually we expect that another field strength defined by 
$\widetilde{F}_{p+2}=e^{\hphi}\,dC_{p+1}$ corresponds to 
an R-R vertex operator of NSR strings in a flat background.
To see this in our formulation, 
we first recall that we have introduced the $(10-d)$-dimensional dilaton 
$\phi$ as a singlet of $O(d,d;\bZ)$. 
This implies that the 10-dimensional dilaton $\hphi$ 
should transform as $e^{\hphi}\propto G^{1/4}$. 
On the other hand, we could have further decomposed the operator 
$\bL_G$ as $\bL_G=\bL_E^{\dag} \bL_E$ where $E=(E_{ia})$ $(i,a=1,...,d)$ 
is a vielbein for $G$, $G=E\,E^T$. 
Then one might say that the state $\bL_E\ket{dC}$ 
corresponds to an R-R vertex in a flat background. 
Thus, noticing that the operator $\bL_E$ will carry 
the factor $G^{1/4}$, 
we expect that $e^{\hphi}\,dC$ will transform as in a flat case.

%%%%%%%%%%%%%%%%%%%%%%%%%%%%%%%%%%%%%%%%%%%%%%%%%%%%%%%%%%%%%%%
%
\section*{Acknowledgment}

The authors would like to thank C.\ Bachas, Y.\ Imamura, K.\ Ito, 
H.\ Kunitomo, S.\ Sugimoto and especially S.\ Mizoguchi 
for useful discussions. 
One of us (T.O.) would also like to thank the Yukawa Institute 
for hospitality. 
The work of M.F.\ is supported in part by the Grand-in-Aid 
for Scientific Research from the Ministry of Education, Science 
and Culture, 
and the work of T.O.\ and the work of H.T.\ are supported in part 
by the JSPS Research Fellowships for Young Scientists.  

%%%%%%%%%%%%%%%%%%%%%%%%%%%%%%%%%%%%%%%%%%%%%%%%%%%%%%%%%%%%%%%
%
~\\
\noindent{\large{\bf Note added}}

\noindent
After the first version of the present paper was put 
on the bulletin board, 
some related works appeared \cite{Ha,CLPS}, 
which also investigate the T-duality 
transformation of R-R fields from a different point of view.

%%%%%%%%%%%%%%%%%%%%%%%%%%%%%%%%%%%%%%%%%%%%%%%%%%%%%%%%%%%%%%%
%
\section*{Appendix: ``Self-dual'' formulation of type II 
effective actions}
\renewcommand{\theequation}{A.\arabic{equation}}

In this appendix, we prove that the original R-R action plus 
the Chern-Simon term, (\ref{ramond})--(\ref{cs}):
\ba
  S^{({\rm IIA})}_{\rm R}+S^{({\rm IIA})}_{\rm CS}
   &=&-\,\frac{1}{4\kappa_{10}^2}\int d^{10}x\sqrt{-\widehat{g}}
    \,\left(\,|\hF_{2}|^2_{\hg}\,+\,|\hF_{4}|^2_{\hg}\,\right)\nn
  &&~~~+\frac{1}{4\kappa_{10}^2}\,\int\,
   \hB_2\we d C_3 \we d C_3 \label{action1}\\
  S^{({\rm IIB})}_{\rm R}+S^{({\rm IIB})}_{\rm CS}
   &=&-\,\frac{1}{4\kappa_{10}^2}\int d^{10}x\sqrt{-\widehat{g}}
    \,\left(\,|\hF_{1}|^2_{\hg}\,+\,|\hF_{3}|^2_{\hg}
      \,+\,\frac{1}{2}\,|\hF_{5}|^2_{\hg}\,\right)\nn
  &&~~~+\frac{1}{4\kappa_{10}^2}\,\int\,
   \hB_2\we d C_4 \we d C_2\n
\ea
with $C_1,C_3$ (or $\RR_1,\RR_3$) and $C_0,C_2,C_4$ 
(or $\RR_0,\RR_2,\RR_4$) being independent variables, 
respectively, is equivalent to the new action (\ref{newramond}):
\ba
  S^{({\rm IIA})}_{\rm R+CS}&\equiv& \frac{1}{8\kappa_{10}^2}\,
   \int\,
   \sum_{p+2=2,4,6,8}\,F_{p+2}\we \ast F_{p+2}
   =-\,\frac{1}{8\kappa_{10}^2}\,
   \int\,\sum_{p+2=2,4,6,8}\,\left|\,F_{p+2}\,\right|_{\hg}^2 
   \label{action2}\\
  S^{({\rm IIB})}_{\rm R+CS}&\equiv&\frac{1}{8\kappa_{10}^2}\,
   \int\,
   \sum_{p+2=1,3,5,7,9}\,F_{p+2}\we \ast F_{p+2}
   =-\,\frac{1}{8\kappa_{10}^2}\,
   \int\,\sum_{p+2=1,3,5,7,9}\,\left|\,F_{p+2}\,\right|_{\hg}^2 \n
\ea
with $\RR_1,\RR_3,\RR_5,\RR_7$ 
and $\RR_0,\RR_2,\RR_4,\RR_6,\RR_8$ being 
independent variables, 
in the sense that both give the same equations of motion 
when the constraints (\ref{elemag}): 
\ba
  \begin{array}{lll}
    \ast F_1=F_9, & \quad &\ast F_2=-F_8, \\
    \ast F_3=-F_7, & \quad & \ast F_4=F_6, \\
    \ast F_5=F_5, & \quad & \ast F_6=-F_4, \\
    \ast F_7=-F_3, & \quad & \ast F_8=F_2, \\
    \ast F_9=F_1 & \quad &
  \end{array}\label{elemag2}
\ea
are imposed on the extra variables, $\RR_5,...,\RR_8$, 
after the equations of motion are derived from (\ref{action2}). 
Here their field strengths are defined by 
\ba
  {}F\equiv\sum_{p+2=1}^9\,F_{p+2}\equiv e^{-\hB_2}\we d\RR
  \label{strength}
\ea
with 
\ba
  \RR\equiv \sum_{p+1=0}^8\,\RR_{p+1},
\ea
and the 10-dimensional Hodge-star $\ast$ is defined by 
\ba
  &&\ast\left( dx^{\hmu_1}\we\cdots\we dx^{\hmu_n}\right)\\
  &&~~~~~~ \equiv\frac{1}{(10-n)!}\,\frac{1}{\sqrt{-\hg}}\,
   \epsilon^{\hmu_1\cdots\hmu_n\hnu_1\cdots\hnu_{10-n}}
   \hg_{\hnu_1\hat{\lambda}_1}...
   \hg_{\hnu_{10-n}\hat{\lambda}_{10-n}}\,
   dx^{\hat{\lambda}_1}\we\cdots\we dx^{\hat{\lambda}_{10-n}}\n
\ea
with $\epsilon^{01\cdots 9}=+1$.
Note that $K$-forms satisfy  
$\ast^2\, \Omega_K=(-1)^{K+1}\,\Omega_K$ and 
\ba
  d^{10}x\,\sqrt{-\hg}\,\left|\,\Omega_K\,\right|^2_{\hg}
   &\equiv& d^{10}x\,\sqrt{-\hg}\,\,\frac{1}{K!}\,\,
    \hg^{\hmu_1\hnu_1}\cdots\hg^{\hmu_K\hnu_K}\,
    \Omega_{\hmu_1...\hmu_K}\,\Omega_{\hnu_1...\hnu_K}\\
  &=& - \Omega_K\we\ast \Omega_K\n
\ea
in 10-dimensional Minkowski space.

{}First, we note that the original action (\ref{action1})
can be written as 
\ba
  S^{({\rm IIA})}_{\rm R}+S^{({\rm IIA})}_{\rm CS}&=&
   \frac{1}{4\kappa_{10}^2}\,\int\,\left(
  {}F_2\we \ast F_2 + F_4 \we\ast F_4
   + \hB_2\, F_4^2+\hB_2^2\, F_4\, F_2 
   + \frac{1}{3}\,\hB_2^3\, F_2^2\right) \nn
  S^{({\rm IIB})}_{\rm R}+S^{({\rm IIB})}_{\rm CS}&=&
   \frac{1}{4\kappa_{10}^2}\,\int\,\left(
  {}F_1\we \ast F_1 + F_3 \we\ast F_3 + \frac{1}{2}\, F_5\we\ast F_5 
   \right. \label{action1a} \\
   &&~~~~~~~~~~~~~~~
    \left. +\, \hB_2\, F_5\, F_3+\frac{1}{2}\,\hB_2^2\, F_5\, F_1 
    + \frac{1}{6}\,\hB_2^3\, F_3 \, F_1\right). \n
\ea
Combined with the NS-NS action $S_{\rm NS}$, (\ref{ns}), 
the equations of motion are thus 

\noindent\underline{IIA}
\ba
  0&=&d\left(\ast F_4 + \hB_2\,F_4 + \frac{1}{2}\,\hB_2^2\,F_2 
      \right) \nn
  0&=&d\left(-\ast F_2 + \hB_2 \ast F_4 + \frac{1}{2}\,\hB_2^2\,F_4
   + \frac{1}{6}\,\hB_2^3\,F_2\right) \label{EOMA}\\
  0&=&d\left(e^{-2\hphi}\,\ast \hB_2\right)+F_2\,\ast F_4
   -\frac{1}{2}\,F_4^2 \n
\ea
\noindent\underline{IIB}
\ba
  0&=&d\left(\ast F_5 + \hB_2\,F_3 + \frac{1}{2}\,\hB_2^2\,F_1 
      \right)\nn
  0&=&d\left(-\ast F_3 + \hB_2\,\ast F_5 + \frac{1}{2}\,\hB_2^2\,F_3
   + \frac{1}{6}\,\hB_2^3\,F_1\right) \label{EOMB}\\ 
  0&=&d\left(\ast F_1 - \hB_2\,\ast F_3 
   + \frac{1}{4}\,\hB_2^2 \left(F_5 + \ast F_5\right)
   + \frac{1}{6}\,\hB_2^3\,F_3 + \frac{1}{24}\,\hB_2^4\,F_1 \right)
      \nn
  0&=&d\left(e^{-2\hphi}\,\ast \hB_2\right)+F_1\,\ast F_3
   +\frac{1}{2}\,F_3\,F_5+\frac{1}{2}\,F_3\,\ast F_5, \n
\ea
as well as the Einstein equation with the energy-momentum tensor 
of R-R fields:
\ba
  T^{({\rm R})}_{\hmu\hnu}&=&\left\{\begin{array}{ll}
   \displaystyle{\cE_{\hmu\hnu}(F_2)+\cE_{\hmu\hnu}(F_4)} 
     & ({\rm IIA})\label{EM}\\
   \displaystyle{\cE_{\hmu\hnu}(F_1)+\cE_{\hmu\hnu}(F_3)
      +\frac{1}{2}\,\cE_{\hmu\hnu}(F_5)} 
     & ({\rm IIB})
   \end{array}\right.
\ea
where $\cE_{\hmu\hnu}$ is defined for an $n$-form 
$F_n=(1/n!)\,F_{\hmu_1...\hmu_n}dx^{\hmu_1}\!\we\!\cdots\!\we\!
dx^{\hmu_n}$ as 
\ba
  \cE_{\hmu\hnu}(F_n)&\equiv&
   \frac{1}{(n-1)!}\,F_{\hmu\,\hmu_1\cdots\hmu_{n\!-\!1}}\,
   {}F_\hnu^{\,\,\,\hmu_1\cdots\hmu_{n\!-\!1}}
   -\frac{1}{2}\,\hg_{\hmu\hnu}\,\left|\,F_n\,\right|_{\hg}^2.
\ea
Equations (\ref{EOMA}) and (\ref{EOMB}) imply that 
$F_1,...,F_5$ can be expressed in the following form 
with integration ``constants'' $\RR_{p+1}$ $(p+1\geq 5)$: 

\noindent\underline{IIA}
\ba
 \ast F_2&=&-\left(e^{-\hB_2}\we d\RR\right)_8 \equiv -F_8\\
 \ast F_4&=& \left(e^{-\hB_2}\we d\RR\right)_6 \equiv F_6\n
\ea
\noindent\underline{IIB}
\ba
 \ast F_1&=&\left(e^{-\hB_2}\we d\RR\right)_9 \equiv F_9\nn
 \ast F_3&=&-\left(e^{-\hB_2}\we d\RR\right)_7 \equiv - F_7\\
 \ast F_5&=&\left(e^{-\hB_2}\we d\RR\right)_5 \equiv F_5 .\n
\ea
{}For example, the first equation of (\ref{EOMA}) is solved as 
\ba
  \ast F_4 + \hB_2 \,F_4 + \frac{1}{2}\,\hB_2^2\,F_2 = d\RR_5
\ea
with some 5-form $\RR_5$. 
Then $\ast F_4$ is written as
\ba
  \ast F_4 &=& d\RR_5-\hB_2 F_4-\frac{1}{2}\,\hB_2^2\,F_2 \nn
  &=&  d\RR_5-\hB_2 (d\RR_3-\hB_2 d\RR_1)
   -\frac{1}{2}\,\hB_2^2\,d\RR_1 \\
  &=& d\RR_5-\hB_2 d\RR_3 +\frac{1}{2}\,\hB_2^2\,d\RR_1 \nn
  &=& \left( e^{-\hB_2}\we d\RR \right)_6 \nn
  &\equiv& F_6.\n
\ea

Now we restart the argument in the reverse order, 
and this time we treat all the fields 
$\RR_{p+1}$ $(p\!+\!1=0,...,8)$ as independent variables 
with field strengths (\ref{strength}),
and adopt (\ref{action2}) plus $S_{\rm NS}$ 
as their action functional. 
The variation of the action with respect to these fields 
can be easily found to be 

\noindent\underline{IIA}
\ba
  0&=&d\left(\ast F_8\right)\nn
  0&=&d\left(-\ast F_6+\hB_2 \ast F_8\right)\nn
  0&=&d\left(\ast F_4-\hB_2 \ast F_6
    +\frac{1}{2}\,\hB_2^2 \ast F_8\right)\\
  0&=&d\left(-\ast F_2+\hB_2 \ast F_4
    -\frac{1}{2}\,\hB_2^2 \ast F_6
    +\frac{1}{6}\,\hB_2^3 \ast F_8\right),\n
\ea
\noindent\underline{IIB}
\ba
  0&=&d\left(\ast F_9\right)\nn
  0&=&d\left(-\ast F_7+\hB_2 \ast F_9\right)\nn
  0&=&d\left(\ast F_5-\hB_2 \ast F_7
    +\frac{1}{2}\,\hB_2^2 \ast F_9\right)\\
  0&=&d\left(-\ast F_3+\hB_2 \ast F_5
    -\frac{1}{2}\,\hB_2^2 \ast F_7
    +\frac{1}{6}\,\hB_2^3 \ast F_9\right)\nn
  0&=&d\left(\ast F_1-\hB_2 \ast F_3
    +\frac{1}{2}\,\hB_2^2 \ast F_5
    -\frac{1}{6}\,\hB_2^3 \ast F_7
    +\frac{1}{24}\,\hB_2^4 \ast F_9
    \right).\n
\ea
These are nothing but the set of the Bianchi identities 
and the equations of motion for the original fields 
$\RR_0,...,\RR_4$ 
if we identify $\ast F_{p+2}=\pm F_{8-p}$ as in (\ref{elemag2}).
{}Furthermore, the variation with respect to $\hB_2$ gives 

\noindent\underline{IIA}
\ba
 0=d\left(e^{-2\hphi}\,\ast \hB_2\right)
   +\frac{1}{2}\,F_2\,\ast F_4+\frac{1}{2}\,F_4\,\ast F_6
   +\frac{1}{2}\,F_6\,\ast F_8 
\ea
\noindent\underline{IIB}
\ba
 0=d\left(e^{-2\hphi}\,\ast \hB_2\right)
   +\frac{1}{2}\,F_1\,\ast F_3+\frac{1}{2}\,F_3\,\ast F_5
   +\frac{1}{2}\,F_5\,\ast F_7 +\frac{1}{2}\,F_7\,\ast F_9, 
\ea
which equal the last equations of (\ref{EOMA}) and (\ref{EOMB}), 
respectively, after the identification (\ref{elemag2}) is made.

The Einstein equation will be accompanied 
by the new energy-momentum tensor for R-R fields:
\ba
  T^{({\rm R})}_{\hmu\hnu}&=&\left\{\begin{array}{ll}
   \displaystyle{\frac{1}{2}\,\sum_{n=2,4,6,8}\,
     \cE_{\hmu\hnu}(F_n)} & ({\rm IIA})\\
   \displaystyle{\frac{1}{2}\,\sum_{n=1,3,5,7,9}\,
     \cE_{\hmu\hnu}(F_n)} & ({\rm IIB})
   \end{array}\right.
\ea
This agrees with the previous one (\ref{EM})
since the following identity holds for 
the dual field $\widetilde{F}_{10-n}\equiv \ast F_n$:
\ba
  \cE_{\hmu\hnu}(\widetilde{F}_{10-n})=\cE_{\hmu\hnu}(F_n),
\ea
which can be easily proved by using 
\ba
  \frac{1}{(9-n)!}\,
   \widetilde{F}_{\hmu\,\hat{\lambda}_1\cdots
     \hat{\lambda}_{9\!-\!n}}\,
   \widetilde{F}_{\hnu}^{\,\,\,\,\hat{\lambda}_1\cdots
     \hat{\lambda}_{9\!-\!n}}
    &=& \frac{1}{(n-1)!}\,
   {}F_{\hmu\,\hat{\lambda}_1\cdots
     \hat{\lambda}_{n\!-\!1}}\,
   {}F_{\hnu}^{\,\,\,\,\hat{\lambda}_1\cdots
      \hat{\lambda}_{n\!-\!1}}
   - \hg_{\hmu\hnu}\,\left|\,F_n\,\right|_{\hg}^2 \, ,\nn
 \left|\,\widetilde{F}_{10-n}\,\right|_{\hg}^2 
   &=& -\,\left|\,F_n\,\right|_{\hg}^2 .
\ea
Since the equivalence for the variation with respect to 
the dilaton $\hphi$ is obvious, 
we have completed the proof of the equivalence 
between the two actions (\ref{action1}) and (\ref{action2}).

%%%%%%%%%%%%%%%%%%%%%%%%%%%%%%%%%%%%%%%%%%%%%%%%%%%%%%%%%%%%%%%
%

\end{document}